\documentstyle[emulateapj,psfig]{article}
\received{25 July 2000}  
\def\etal{{\it et al.}}
\def\lsim{\hbox{ \rlap{\raise 0.425ex\hbox{$<$}}\lower 0.65ex\hbox{$\sim$} }}
\def\gsim{\hbox{ \rlap{\raise 0.425ex\hbox{$>$}}\lower 0.65ex\hbox{$\sim$} }}

\def\arcsec{\hbox{$^{\prime\prime}$}}

\def\f(h{\hbox{$~\!\!^{\rm h}$}}

\def\ale{\mathrel{\hbox{\rlap{\hbox{\lower4pt\hbox{$\sim$}}}\hbox{$<$}}}}
\def\age{\mathrel{\hbox{\rlap{\hbox{\lower4pt\hbox{$\sim$}}}\hbox{$>$}}}}

\slugcomment{Submitted to The Astrophysical Journal Letters}

\lefthead{Djorgovski, Bloom, \& Kulkarni} 
\righthead{The Redshift and Host Galaxy of GRB 980613}

\begin{document}

\title{The Redshift and the Host Galaxy of GRB 980613:\\
A Gamma-Ray Burst From a Merger-Induced Starburst?
\footnotemark}

\footnotetext{Partially based on the observations obtained at the
W.~M.~Keck Observatory which is operated by the California Association
for Research in Astronomy, a scientific partnership among California
Institute of Technology, the University of California and the National
Aeronautics and Space Administration.}

\author{S. G. Djorgovski, J. S. Bloom, S. R. Kulkarni}

\affil{Palomar Observatory 105--24, California Institute of Technology,
            Pasadena, CA 91125, USA; {\tt george,jsb,srk@astro.caltech.edu}}

%%%%%%%%%%%%%%%%%%%%%%%%%%%%%%%%%%%%%%%%%%%%%%%%%%%%%%%%%%%%%%%%%%%%%%%%%%%%%
\begin{abstract}
We present optical and near-IR identification and spectroscopy of the
host galaxy of GRB 980613.  The burst was apparently associated with
the optically (restframe UV) brightest component of a system of at
least five galaxies or galaxy fragments at a redshift of $z = 1.0969$.
The component we identify as the host galaxy shows a moderately high
unobscured star formation rate, SFR $\sim 5~M_\odot$ yr$^{-1}$, but a
high SFR per unit mass, indicative of a starburst.  The image
components show a broad range of $(R-K)$ colors, with two of them
being very red, possibly due to dust.  Overall morphology of the
system can be naturally interpreted as a strong tidal interaction of
two or more galaxies, at a redshift where such events were much more
common than now.  Given the well established causal link between
galaxy mergers and starbursts, we propose that this is a strong case
for a GRB originating from a merger-induced starburst system.  This
supports the proposed link between GRBs and massive star formation.

\end{abstract}


\section{Introduction}

Studies of the cosmic gamma-ray bursts (GRBs) are currently one of the
most active and exciting fields of astrophysics.  A deluge of
important clues in the long-standing puzzle of GRBs began with the
discovery of long-lived X-ray afterglows by the BeppoSAX satellite
(Costa \etal\ 1997\nocite{cfh+97}).  This was followed by the
discovery of optical (\cite{vgg+97}) and radio afterglows (Frail
\etal\ 1997\nocite{fkn+97}), and then the first redshift measurement
which demonstrated the cosmological nature of GRBs (Metzger \etal\
1997).  To date, about a dozen redshift measurements of GRBs have been
obtained.  Studies of afterglows confirmed the synchrotron shock model
({\it e.g.},~\cite{wrm97}), and their physics now seems to be
reasonably well understood.  For a recent review of observational
results, see, {\it e.g.}, Kulkarni \etal\ (2000)\nocite{kbb+00}.

What is still not known is what are both the nature of the progenitors
and the trigger that causes GRBs.  The two currently popular models
include mergers of neutron stars or other compact stellar remnants,
leading to a creation of a black hole, and a collapsar or hypernova
model in which an explosion of a massive star produces a black hole
remnant.  In both cases, spin energy of a debris torus or of the black
hole itself is used to power the GRB.  While other models are still
viable, there is now a growing evidence in favor of the
collapsar/hypernova model, at least for the long-duration bursts
detected by BeppoSAX. The most direct evidence favoring this was the
probable detection of supernov\ae\ associated with GRB 980326
(\cite{bdk+99}) and GRB 970228 (\cite{rei99,gtv+00}), as well as the
still controversial case of GRB 980425 and SN 1998bw
(\cite{gvv+98,kfw+98}).  All models involving massive stars and their
remnants suggest that GRBs should be closely related to the massive
star formation in galaxies.

Study of GRB host galaxies can provide valuable clues which can
constrain the models, and redshifts are necessary in order to derive
the physical parameters of the GRBs itself, primarily the energy
scale, as well as the hosts: their luminosities, star formation rates,
morphology, locations of GRBs within them, etc.

In this $Letter$ we present deep imaging and spectroscopic
observations of the host galaxy of GRB 980613.  GRB 980613 was
detected and localized by BeppoSAX (\cite{pc+98,pir98a}).  Following
several unsuccessful attempts an optical transient (OT) was discovered
by Hjorth \etal\ (1998)\nocite{hap+98}.  The detection of the host
galaxy was reported by Djorgovski \etal\ (1998a)\nocite{dkoe98}.  Its
redshift determination was first reported by Djorgovski \etal\
(1999)\nocite{dkb+98a}, and is described in more detail here, with
additional data.

\section{Observations and Data Reductions}
\label{sec:obs}

In what follows we assume the Galactic foreground extinction of $A_V =
0.27$ mag (\cite{sfd98}), and use the Galactic extinction curve from
Cardelli, Clayton, \& {Mathis} (1998)\nocite{ccm98} with $R =
A_V/E_{B-V} = 3.1$.

Our initial imaging date were obtained on the W.~M.~Keck Observatory
10-m telescope (Keck II) by Dr.~H.~Ebeling, on 16.30 June 1998 UT, in
the $R$ band, using the Low-Resolution Imaging Spectrometer (LRIS;
\cite{occ+95}); two 300 s exposures were obtained.  We reduced these
images in the standard manner and confirmed that no cosmic rays
affected the immediate area of the transient.  After the transient had
faded we reobserved the field in $R$ band (29 November 1998 UT; 900s),
$I$ band (24 March 1999 UT; 1000s), and $K$ band (7 February 1999 UT;
2040s).  The final combined images in each filter is shown in fig.~1.
The morphology of the system surrounding the GRB is complex.  We label
5 distinct components in fig.~2. Table 1 provides their magnitudes and
offsets relative to the position of the OT.
\begin{figure*}[tbp]
\centerline{\psfig{file=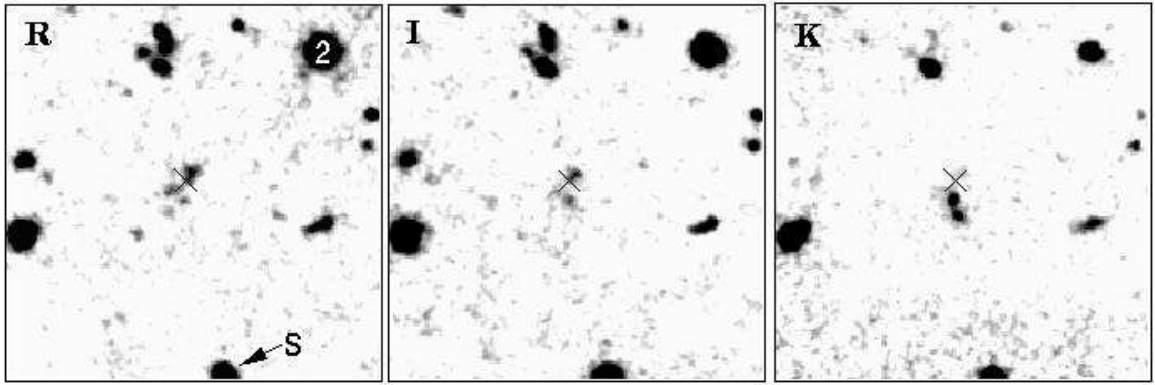,width=6.2in,angle=270}}
\caption[]{Late time images of the field of GRB 980613 with the Keck
Telescopes in $R$ (left), $I$ (middle) and $K$ bands (right).  Field
size is 37\arcsec $\times$ 37\arcsec, with N up and E to the left.
The ``$\times$'' marks the location of the OT.  The offset star used
for our spectrosopic observations is labeled ``S'' (the OT was
21.0\arcsec N and 3.9\arcsec E from the star), as is the Halpern et
al.~(1998) star ``2'' used for photometric zeropointing of our $R$ and
$I$ band images. }
\label{fig:0613rik}
\end{figure*}

We referenced our $R$ and $I$ band magnitudes using star 2 (Halpern
\etal\ 1998\nocite{hf98a}; Diercks \etal\ 1998\nocite{dds+98}).  For
each component we used a circular aperature and
performed an aperature correction using the curve-of-growth of star 2.
The total magnitude of system A+B+C is $R=23.11 \pm 0.05$.  We
estimate the zeropoint systematic uncertainty due to an uncertain
color correction at 0.15 mag.  The 2-$\sigma$ upper limit to a point
source detection is $R = 25.9$ mag and $I = 24.9$ mag.  Our $K$ band
imaging was taken under photometric conditions and we used SJ 9134
(Persson \etal\ 1998\nocite{pmk+98}) to obtain a zeropoint for the
night.  The $K$ band aperature magnitudes of the 5 components are listed
in Table 1.

We registered the images from 16 June 1998 and 29 November 1998 $R$
band images using 7 unsaturated stellar objects common to both images
within 30 arcsec from the host galaxy.  We used the optimum-filter
technique in the IRAF package CENTER to position the astrometric tie
objects.  We determined the rotation, shift, and relative scale of the
two images using GEOMAP and then registered the early-time image to
the late-time image.  The peak of the OT and the putative host were
then estimated.  Including the peak center errors and the registration
uncertainty we find the OT was offset from the brightest $R$ band peak
(A) by 0.52\arcsec $\pm$ 0.13\arcsec\ East and 0.83\arcsec $\pm$
0.14\arcsec S.  This amounts to a radial angular projected offset of
0.98\arcsec $\pm$ 0.14\arcsec, and we interpret component A as the
most likely host galaxy of the GRB.  The $I$ and $K$ band images of
the host were also registered to the late-time $R$ band image. Fig.~2
shows a color composite 15.3\arcsec $\times$ 15.3\arcsec region around
the GRB and its host using the late-time $R$, $I$, and $K$ band
images.  The ellipse is a 3-$\sigma$ error contour about the position
of the GRB.

Spectra of the host galaxy were obtained using LRIS on 17 December
1998, 14 January 1999, and 16 February 1999 UT, all in good observing
conditions.  We used a 1.5 arcsec wide long slit, always at a position
angle close to the parallactic.  On December 14, we used a 300 lines
mm$^{-1}$ grating, giving an effective instrumental resolution FWHM
$\approx 16\,$\AA and an approximate wavelength coverage 3950--8950
\AA, and obtained 6 exposures totaling 8300 s.  On January 14 and
February 16, we used a 600 lines mm$^{-1}$ grating, giving an
effective instrumental resolution FWHM $\approx 8$ \AA and an
approximate wavelength coverage 5970--8540 and 5700--$8270\,$\AA,
respectively; 3 exposures of 1800 s were obtained on each night.  The
object was dithered on the spectrograph slit by several arcsec between
the exposures.  Exposures of an internal flat-field lamp and arc lamps
were obtained at comparable telescope pointings immediately following
the target observations.  Exposures of standard stars from Oke \& Gunn
(1983)\nocite{og83} and Massey \etal\ (1998)\nocite{msb88} were
obtained and used to measure the instrument response curve.  We
estimate the net flux zero-point uncertainty, including the slit
losses, to be about 10--20\%.

\vskip 0.2cm
\begin{minipage}[b]{8.9cm}
\centerline{\psfig{file=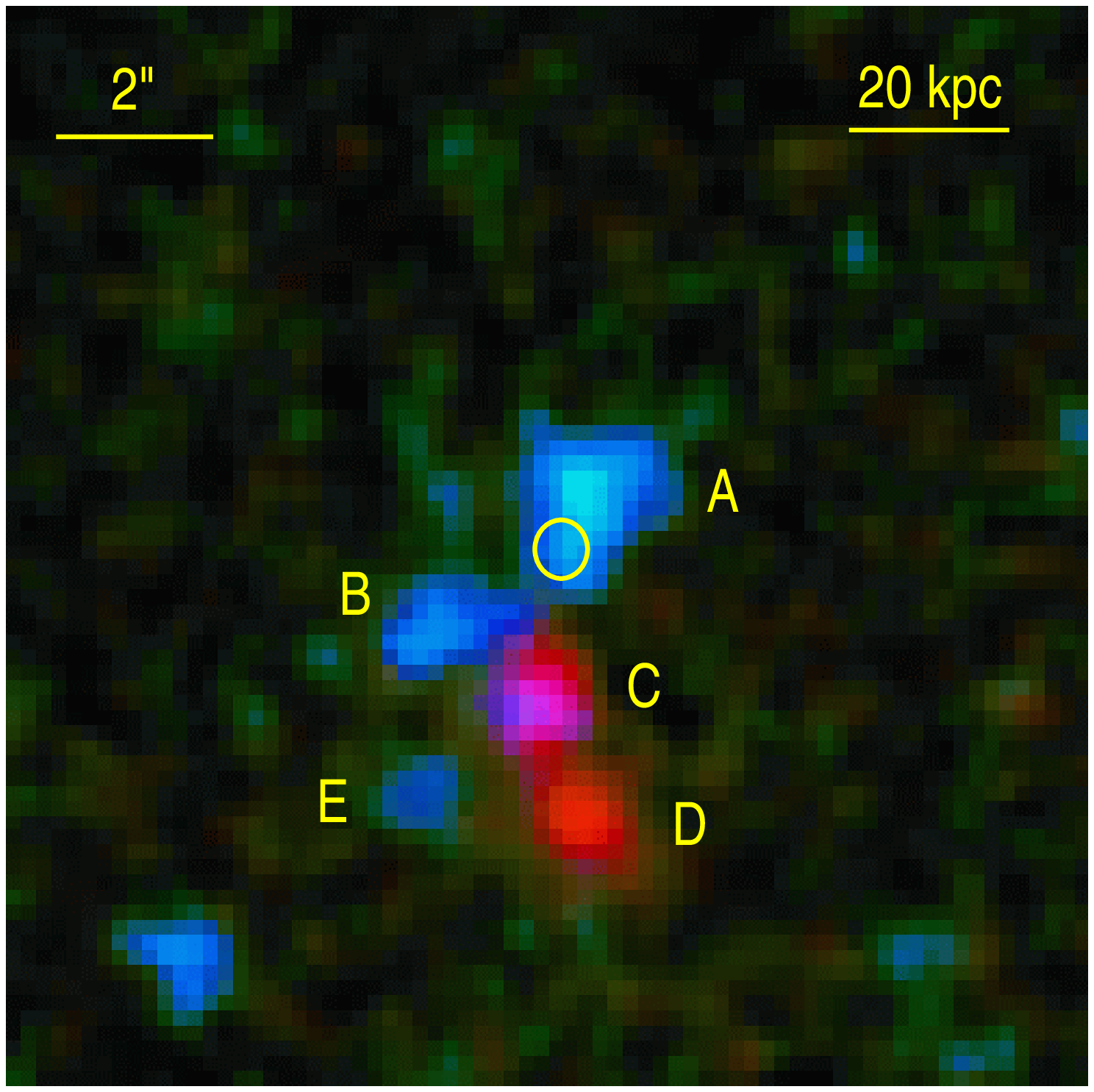,width=3.2in,angle=270}} 
\vskip -0.2cm
\figcaption[]{Color composite 15.3\arcsec\ $\times$ 15.3\arcsec\ region
around the GRB and its host. We have used the late-time $R$, $I$, and
$K$ band images for the red, green, and blue channels. Before combining,
we subtracted the mode-determined sky level and scaled each image by
the r.m.s.~noise of the background. The ellipse is a 3-$\sigma$ error
contour about the position of the GRB.  The five distinct components
A---E are labeled. The physical scale was is computed assuming all
objects are in a redshift sheet at $z=1.096$. As discussed in the
text, the GRB appears within the light of component ``A'', which we interpret
as the host galaxy. }
\label{fig:0613color}
\end{minipage}

Wavelength solutions were obtained from arc lamps in the standard
manner, and then a second-order correction was determined from the
wavelengths of isolated strong night sky lines, and applied to the
wavelength solutions.  This procedure largely eliminates systematic
errors due to the instrument flexure, and is necessary in order to
combine the data obtained during separate nights.  The final
wavelength calibrations have the r.m.s.~$\sim 0.2 - 0.5$ \AA, as
determined from the scatter of the night sky line centers.  All
spectra were then rebinned to a common wavelength scale with a
sampling of 2.5 \AA\ using a Gaussian with a $\sigma = 2.5$ \AA\ as
the interpolating/weighting function.  This is effectively a very
conservative smoothing of the spectrum, since it is smaller than the
instrumental resolution.  Individual spectra were extracted and
combined using a statistical weighting based on the signal-to-noise
ratio determined from the data themselves (rather than by the exposure
time).
\begin{figure*}[tbp]
\centerline{\psfig{file=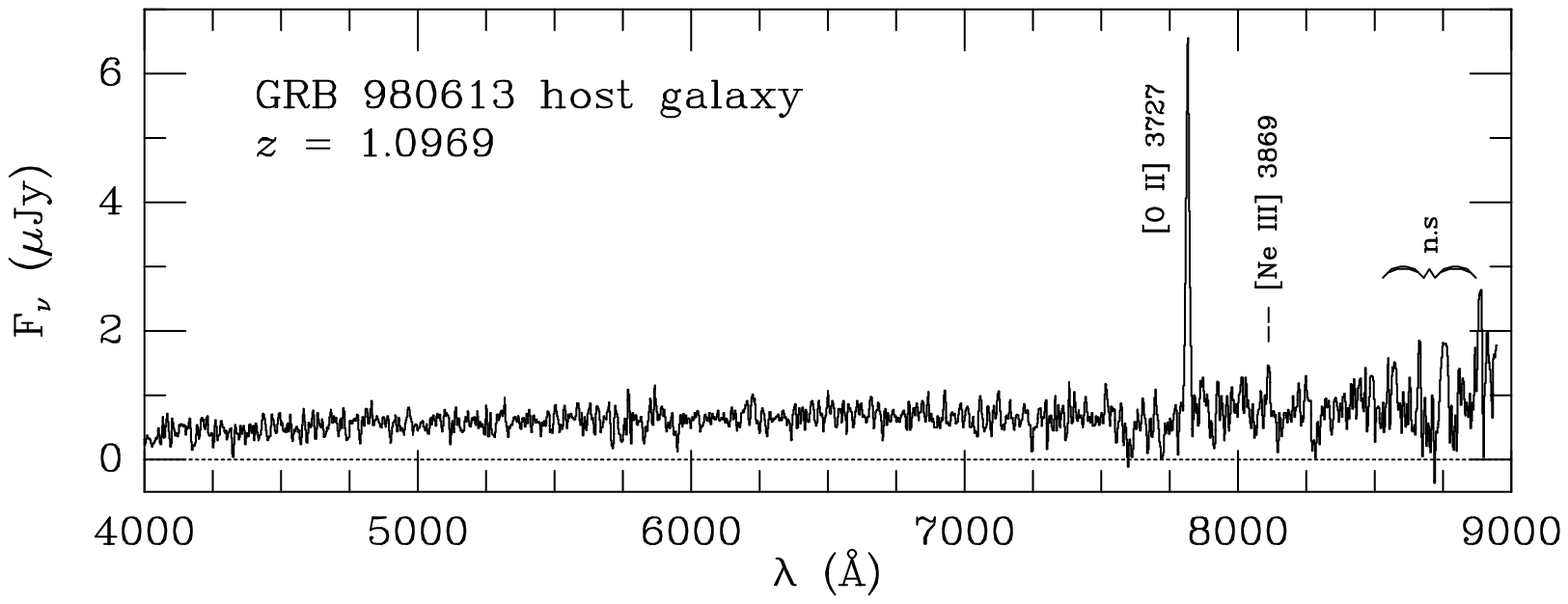,width=6.2in}}
\caption[]{The weighted average spectrum of the host galaxy of GRB
980613 (component A in fig.~2), obtained at the Keck II telescope.
Emission lines [O II] 3727 and [Ne III] 3869 are labeled; ``ns''
refers to noise spikes from a poor subtraction of strong night sky
lines.  }
\label{fig:0613spec}
\end{figure*}

The final combined spectrum of the galaxy is shown in fig.~3.  A
strong [O II] 3727 line emission is present, and a weaker [Ne III]
3869 line is also seen.  The weighted mean redshift is $z = 1.0969 \pm
0.0002$.  The observed continuum is moderately blue, with the
spectrophotometric colors $(B-V) \approx 0.45$ and $(V-R) \approx
0.35$ mag.

The corrected [O II] 3727 line flux is $(5.25 \pm 0.15) \times
10^{-17}$ erg cm$^{-2}$ s$^{-1}$ Hz$^{-1}$, and its observed
equivalent width is $W_\lambda = 125 \pm 5$ \AA, {\it i.e.}, $60 \pm
2.4$ \AA\ in the restframe, which is on a high side for field galaxies
at comparable redshifts, but not extraordinary (\cite{hcbp98}).  The
[Ne III] 3869 line has a flux of about 10\% of the [O II] 3727 line.
The continuum flux at $\lambda_{obs} = 5871\,$\AA, corresponding to
$\lambda_{rest} = 2800\,$\AA, is $F_\nu = 0.84 ~\mu$Jy, uncertain by a
few percent (statistical).  Additional spectrophotometric flux
zero-point uncertainty is estimated to be $\sim 10-20$\%.

The continuum flux from our direct photometry in the $I$ band, which
corresponds roughly to the restframe $B$ band, is $F_\nu = 1.15
~\mu$Jy, uncertain by $\sim 10$\%.

\section{Discussion}
\label{sec:0613imp}

We will assume a flat cosmology with $H_0 = 65$ km s$^{-1}$
Mpc$^{-1}$, $\Omega_M = 0.3$, and $\Lambda_0 = 0.7$.  For $z =
1.0969$, the luminosity distance is $2.461 \times 10^{28}$ cm, and 1
arcsec corresponds to 8.8 proper kpc in projection.

The gamma-ray fluence from this burst was $(1.71 \pm 0.25) \times
10^{-6}$ erg cm$^{}-2$ (\cite{wkc+98}).  The corresponding isotropic
$\gamma$-ray energy is $E_{\gamma,iso} = 6.2\times 10^{51}$ erg.

From the [O II] 3727 line flux, we derive the line luminosity
$L_{3727} = 4.0 \times 10^{41}$ erg s$^{-1}$.  Using the star
formation rate estimator from Kennicutt (1998)\nocite{ken98}, we
derive the SFR $\approx 5.6 ~M_\odot$ yr$^{-1}$.  From the UV
continuum luminosity at $\lambda_{rest} = 2800$\AA, following Madau,
Pozzetti \& Dickinson (1998)\nocite{mpd98}, we derive SFR $\approx 3.1
~M_\odot$ yr$^{-1}$.  The difference may be due to the internal
reddening within the host galaxy, but we also note that the [O II]
line estimator is more sensitive to the current or recent massive star
formation than the UV continuum estimator.  This is consistent with
the presence of the [Ne III] 3869 line, also seen in spectra of some
other GRB hosts (\cite{bdkf98,bdk01}) which may be indicative of a
recent, very massive star formation.  We are completely insensitive to
any fully obscured star formation component, so that these numbers
represent lower limits.  The only GRB host galaxy with a higher
unobscured SFR measured to date is the host of GRB 980703
(\cite{dkb+98b}).

From the observed continuum flux in the restframe $B$ band, we derive
$M_B \approx -19.85$ mag, which is about 1 mag fainter than the
present-day $L_*$ galaxy.  Considering that an average galaxy at this
redshift may be $\sim 0.5 - 1$ mag brighter than today due to normal
evolution effects, we conclude that this galaxy (A) is moderately
underluminous.  The star formation rate per unit mass is thus fairly
high, consistent with the large equivalent width of the [O II] line.
We thus conclude that this galaxy is undergoing a starburst, albeit
mild.

However, the most interesting feature may be the morphology of the
entire system (A--E), which is highly suggestive of an interaction or
early stages of a merger of at least two galaxies, some of which may
be partly obscured.  Redshifts of all 5 components are necessary to
really test this interpretation; we note, however, that we detect a
weak [O II] line emission coincident with the component C in our
long-slit spectra, at the same redshift as the component A.  Another
possibility is that we are seeing a chance superposition of galaxies
at very different redshifts, e.g., that the IR-bright components C and
D may be unrelated to A (and perhaps also B and E), and if they are in
the foreground, some gravitational lensing may be involved.  However,
we consider this less likely than the interaction/merger hypothesis.

The rate of galaxy interactions and mergers increases sharply with
redshift ({\it cf.}, \cite{fal+00}), and finding a strongly
interacting system at $z \sim 1.1$ is not surprising.  The projected
separations of components A--E (see fig.~2) are typical of
intergalactic separations where dark halos are overlapping and the
tidal friction inevitably leads to a merger.

The very red colors of the components C and D, $(R-K) \approx 4.4$ and
$> 5.8$, respectively, is naturally explained due to obscuration by
dust, which is consistent with the merger hypothesis.  Alternatively,
they could be passively evolving ellipticals which formed at a very
high redshift, which requires some fine-tuning of model parameters.

It is now well established that mergers of gas-rich galaxies lead to
bursts of star formation.  This may be the strongest case so far for a
GRB from such a system.  The only other published case to date,
showing morphology consistent with a mild tidal interaction is the
host galaxy of GRB 990123 ({\it e.g.}, \cite{bod+99,ftm+99,hh99}).
This provides a further evidence in support of the connection of GRBs
with massive star formation.

\acknowledgments

We want to thank the staff of the W.~M.~Keck Observatory for their
expert assistance, and to Dr. Harald Ebeling for obtaining the early
images of the field.  We thank Dana Sadava for help in $K$ band
imaging reductions and Titus Galama for comments. JSB gratefully
acknowledges the fellowship from the Fannie and John Hertz
Foundation. SGD acknowledges partial funding from the Bressler
Foundation. This work was supported in part by grants from the NSF and
NASA to SRK.

\begin{deluxetable}{lcccc}
\footnotesize
\tablecaption{Photometry and Offsets of Components Near GRB 980613}
\label{tab:0613phot}
\tablewidth{0in} 
\tablehead{ \colhead{Comp.} & \colhead{Offset (\arcsec)}&  
\colhead{R}  &  \colhead{I}  & \colhead{K}  \\
\colhead{}   &\colhead{ w.r.t. OT}       & \colhead{(mag)} & \colhead{(mag)} & \colhead{(mag)}
}\tablecolumns{5} \startdata
A     & 0.52 W, 0.83 N  & $23.81 \pm 0.06$ & $23.42 \pm 0.10$ & $21.65 \pm 0.22$ \\
B     & 1.58 E, 1.18 S  & $24.95 \pm 0.17$ & $24.12 \pm 0.16$ & $> 22.3$         \\
C     & 0.25 E, 2.26 S  & $24.45 \pm 0.10$ & $23.81 \pm 0.12$ & $20.08 \pm 0.06$ \\
D     & 0.47 W, 3.84 S  & $> 25.9$         & $> 24.9$         & $20.12 \pm 0.06$ \\
E     & 1.72 E, 3.44 S  & $25.2 \pm 0.3$   & $24.66 \pm 0.28$ & $> 22.3$       
\enddata
\end{deluxetable}

\end{document}